\documentclass[12pt,a4paper]{article}

\usepackage{amsmath,amssymb}
\usepackage[nosort]{cite}
\usepackage[hyperref,bulletsep]{collect}

\makeatletter \@addtoreset{equation}{section} \makeatother

\makeatletter
\let\old@startsection=\@startsection
\let\oldl@section=\l@section
\renewcommand{\@startsection}[6]{\old@startsection{#1}{#2}{#3}{#4}{#5}{#6\mathversion{bold}}}
\renewcommand{\l@section}[2]{%
\vspace{-0.5em}%
\oldl@section{\mathversion{bold}#1}{#2}}
\makeatother

\makeatletter
\let\old@makecaption=\@makecaption
\def\@makecaption{\small\old@makecaption}
\makeatother

\let\oldPhi=\Phi
\let\oldPsi=\Psi
\let\oldGamma=\Gamma
\let\oldDelta=\Delta
\let\oldSigma=\Sigma
\let\oldTheta=\Theta
\let\oldPi=\Pi
\renewcommand{\Phi}{\mathnormal{\oldPhi}}
\renewcommand{\Psi}{\mathnormal{\oldPsi}}
\renewcommand{\Gamma}{\mathnormal{\oldGamma}}
\renewcommand{\Sigma}{\mathnormal{\oldSigma}}
\renewcommand{\Delta}{\mathnormal{\oldDelta}}
\renewcommand{\Theta}{\mathnormal{\oldTheta}}
\renewcommand{\Pi}{\mathnormal{\oldPi}}

\newcommand{\gen}[1]{\mathfrak{#1}}

\newcommand{\sTr}{\mathop{\mathrm{sTr}}}

\ifx\genfrac\sdflkaj

\else

\fi

\newcommand{\alg}[1]{\mathfrak{#1}}

\newcommand{\comm}[2]{[#1,#2]}

\newcommand{\gcomm}[2]{[#1,#2\}}
\newcommand{\gacomm}[2]{\{#1,#2]}

\newcommand{\state}[1]{\mathopen{|}#1\mathclose{\rangle}}

\newcommand{\nln}{\nonumber\\}

\newcommand{\earel}[1]{\mathrel{}&\hspace{-2\arraycolsep}#1\hspace{-2\arraycolsep}&\mathrel{}}
\newcommand{\eq}{\earel{=}}

\def\[{\begin{equation}}
\def\]{\end{equation}}
\def\<{\begin{eqnarray}}
\def\>{\end{eqnarray}}

\makeatletter
\def\mr@ignsp#1 {\ifx\:#1\@empty\else #1\expandafter\mr@ignsp\fi}%
\newcommand{\multiref}[1]{\begingroup
\xdef\mr@no@sparg{\expandafter\mr@ignsp#1 \: }%
\def\mr@comma{}%
\@for\mr@refs:=\mr@no@sparg\do{\mr@comma\def\mr@comma{,}\ref{\mr@refs}}%
\endgroup}
\makeatother

\newcommand{\hypref}[2]{\ifx\href\asklfhas #2\else\href{#1}{#2}\fi}

\renewcommand{\eqref}[1]{(\multiref{#1})}

\ifx\href\asklfhas\newcommand{\href}[2]{#2}\fi
\newcommand{\arxivno}[1]{\href{http://arxiv.org/abs/#1}{#1}}

\begin{document}

\thispagestyle{empty}
\begin{flushright}\footnotesize
\texttt{\arxivno{hep-th/0610283}}\\
\texttt{PUTP-2213}%
\vspace{0.5cm}
\end{flushright}
\vspace{0.1cm}

\renewcommand{\thefootnote}{\fnsymbol{footnote}}
\setcounter{footnote}{0}

\begin{center}%
{\Large\textbf{\mathversion{bold}%
Yangian Symmetry at Two Loops for \\
the $\alg{su}(2|1)$ Sector of $\mathcal{N}=4$ SYM}\par}
\vspace{2cm}

\textbf{Benjamin I. Zwiebel}
\vspace{5mm}

\textit{
Joseph Henry Laboratories, Princeton University\\
Princeton, NJ 08544, USA \\} 
\vspace{3mm}
\texttt{bzwiebel@princeton.edu}
\vspace{1.8cm}

\textbf{Abstract}\vspace{5mm}

\begin{minipage}{14.8cm}
We present the perturbative Yangian symmetry at next-to-leading order in the $\alg{su}(2|1)$ sector of planar $\mathcal{N}=4$ SYM.  Just like the ordinary symmetry generators, the bi-local Yangian charges receive corrections acting on several neighboring sites. We confirm that the bi-local Yangian charges satisfy the necessary conditions: they transform in the adjoint of $\alg{su}(2|1)$, they commute with the dilatation generator, and they satisfy the Serre relations. This proves that the sector is integrable at two loops.

\end{minipage}

\vspace*{\fill}

\end{center}

\newpage
\setcounter{page}{1}
\renewcommand{\thefootnote}{\arabic{footnote}}
\setcounter{footnote}{0}

\section{Introduction}
The dilatation generator of planar $\mathcal{N} = 4$ SYM is integrable at one-loop \cite{Lipatov:1997vu, Minahan:2002ve, Beisert:2003yb}, and all field theory calculations done so far are consistent with integrability persisting at least through three loops \cite{Staudacher:2004tk}.  Assuming integrability, an asymptotic set of all-loop Bethe equations have been found \cite{Beisert:2005fw}.   Moreover, the dual non-interacting string theory on $AdS_5 \times S^5$ is integrable classically \cite{Bena:2003wd}, and up to a scalar factor, the same Bethe equations give the string quantum spectrum  \cite{Arutyunov:2004vx}, at least up to one loop \cite{Hernandez:2006tk}.  In fact, supersymmetry fixes the string and gauge Bethe equations uniquely, up to a scalar factor \cite{Beisert:2005tm}.  Recent developments on the scalar factor and crossing symmetry \cite{Janik:2006dc} include \cite{Beisert:2006ib, Beisert:2006ez, Bern:2006ew}.

Undoubtedly, the Bethe equations are the most efficient way to obtain the spectrum of the string and gauge theories. However, they depend on the assumption of integrability. Since there is no obvious definition of the R-matrix for this system, perhaps the only way to prove integrability is to construct the Yangian symmetry. The string theory's classical Yangian charges were introduced in \cite{Bena:2003wd}, and \cite{Berkovits:2004xu} argues that an infinite family of non-local charges persists quantum mechanically. For the gauge theory, the Yangian charges are known at leading order for the full theory \cite{Dolan:2003uh, Dolan:2004ps}. For the $\alg{su}(2)$ sector, they are know up to second \cite{Agarwal:2004sz} and fourth order \cite{Beisert:2006xx}, and \cite{Agarwal:2005ed} gave the next-to-leading order Yangian symmetry of the $\alg{su}(1|1)$  sector. By considering the $\alg{su}(2|1)$ sector at next-to-leading order, we encounter important features that have not appeared in these previous studies of the gauge theory Yangian. The Hamiltonian (dilatation generator) is part of the local symmetry algebra, and other local symmetry generators have two-site interactions. 
 
At leading order, the infinite tower of Yangian charges are generated by repeated commutators of the bi-local generators:
\[ 
\gen{Y}^A = \sum_{i < j} f^A{}_{CB} \gen{J}^B(i) \gen{J}^C (j), 
\]
where the $\gen{J}^A$ are local symmetry generators and $f^{A}_{\; \; BC}$ are the structure constants. Here we have introduced the notation $\gen{J}(i)$ to represent the generator acting on site $i$ of the spin chain. 

We propose that the $\mathcal{O}(g^n)$ perturbative corrections to $\gen{Y}^A$, which we write as $\gen{Y}^A_n$, are of the schematic form 
\[
\gen{Y}^A_n = \sum_{\substack{p+q=n \\ i << j}} f^A{}_{CB} \gen{J}^B_p(i) \gen{J}^C_q (j)  + \text{local terms} 
\]
The $<<$ symbol in the summation excludes terms for which the two (generically multi-site) symmetry generators act on common sites, which are instead included in the local terms. Via explicit computation, we confirm that there are Yangian charges of this form for the $\alg{su}(2|1)$ sector at next-to-leading order. Furthermore, in this case the local terms can be expressed simply as sums over products of two overlapping local symmetry generators.

As is well known, for typical integrable systems the Yangian charges only commute with the spin-chain Hamiltonian on infinite length chains. The reason is that it is not possible to consistently define periodic boundary conditions for the Yangian charges.  At next-to-leading order, we find that an infinite chain is required for the Yangian charges to transform properly under local symmetry transformations, to satisfy the Serre relations, and to commute at next-to-leading order (two-loops) with the Hamiltonian\footnote{Note that the leading order Hamiltonian is $\mathcal{O}(g^2)$ or one-loop, so that showing it has $n$th order Yangian symmetry proves $(n+1)$-loop integrability.}.

In typical integrable systems, the existence of Yangian charges follows from a nearest-neighbor R-matrix satisfying the Yang-Baxter equation, which also generates an infinite family of local commuting charges. However, it is unclear if it possible to apply the R-matrix formalism to this system beyond leading order. Therefore, it is reasonable to use the existence of Yangian symmetry as a substitute definition of integrability, and the results of this paper prove that the $\alg{su}(2|1)$ sector is integrable at two-loops. 

In section \ref{algebra} we review the $\alg{su}(2|1)$ algebra and its Yangian generalization, and in section \ref{leadingorder} we review the leading order representation. The next-to-leading order corrected Yangian charges and the proofs that they satisfy the Yangian algebra and commute with the Hamiltonian are given in section \ref{oneloop}. We conclude and give possible directions for future research in section \ref{conclusion}.

\section{The Algebra \label{algebra}}

\subsection{The $\alg{su}(2|1)$ Algebra}

According to our convention, the eight symmetry generators of the $\alg{su}(2|1)$ algebra, $\gen{J}^A, \, A= 1,\ldots 8$, split as follows. The supersymmetry generators have indices between 1 and 4, the bosonic $\alg{su}(2)$ generators have indices between 5 and 7, and  $\gen{J}^8$ is a linear combination of the dilatation generator and the length generator.  These generators are simple linear combinations of those used in \cite{Beisert:2003ys}, where the supersymmetry generators are $\gen{Q}^a$ and $\gen{S}^a$, and the bosonic generators are $\gen{R}^a_b$ and $\gen{D}$ ($a,\,b=1,\,2$). We give these linear combinations in the appendix.
The algebra is 
\<
\gcomm{\gen{J}^A}{\gen{J}^B} = f^{ABC} g_{CD} \gen{J}^D = f^{AB}{}_D \gen{J}^D \label{deff},
\>
where $g$ is the Cartan-Killing form or metric for the $\alg{su}(2|1)$ algebra. The structure constants $f^{ABC}$ are totally antisymmetric, with extra minus signs for every interchange of fermionic indices. Similarly, the metric is symmetric, up to an extra minus sign for switching fermionic indices.  Therefore, we must be careful about the order of the indices of the (inverse) metric when (raising) lowering indices. In our convention, the first index of the (inverse) metric is summed over, as in (\ref{deff}).

We also have the symmetric invariant tensor, defined by 
\<
d^{ABC} = \sTr \gacomm{\gen{J}^A_0}{\gen{J}^B_0} \gen{J}^C_0.
\>
For this definition, all the generators act on the same site, and the mixed brackets with the curly bracket first means that we use the anti-commutator for commuting generators and vice-versa. 

We give the non-vanishing components of the metric, structure constants, and symmetric invariant tensors in the appendix.

\subsection{The $\alg{su}(2|1)$ Yangian algebra}
The $\alg{su}(2|1)$ Yangian Algebra is an infinite-dimensional algebra generated by $\gen{J}^A$ and $\gen{Y}^A$.  Commutators of these generators yield an infinite sequence of generators, $\gen{Y}_{(i)}^A$, ($i = 0, 1, 2, \ldots$). $i=0,1$ correspond to $\gen{J}^A$ and $\gen{Y}^A$. The algebra is defined by the following commutation relations
\< \gcomm{\gen{J}^A}{\gen{J}^B} \eq f^{AB}{}_{C} \gen{J}^C \label{local}, \\
 \gcomm{\gen{J}^A}{\gen{Y}^B} \eq f^{AB}{}_{C} \gen{Y}^C, \label{adjoint} \\
 f^{[BC}{}_{E}\gcomm{\gen{Y}^{A \} }}{\gen{Y}^E}  \eq h^2  (-1)^{(EM)} f^{AK}{}_{D} f^B{}_{E}{}^{L} f^C{}_{F}{}^{M}  f_{KLM}\{ \gen{J}^D, \gen{J}^E, \gen{J}^F \} \label{serre} .
\>
The notation $ \{ \gen{J}^D, \gen{J}^E, \gen{J}^F \} $ denotes the totally symmetric product (with extra minus signs for every interchange of fermionic generators). $(-1)^{(EM)}$ gives $-1$ when both indices are fermionic, and $1$ otherwise. Our conventions lead to $h^2 = -\frac{64}{3}$. The mixed brackets around the raised indices on the left side of the last equation mean that it is summed anti-symmetrically over all permutations of $A,\,B$ and $C$. We obtained (\ref{serre}) (the Serre relation) by substituting (\ref{adjoint}) into the standard form for this Serre relation. Usually this would have no effect, but in our case this simplifies the proof of that the Serre relation is satisfied. There is another Serre relation that Yangian algebras must satisfy. However, this relation follows from the ones given above unless (\ref{serre}) is trivial \cite{Drinfeld:1985rx, Drinfeld:1986in}, and (\ref{serre}) is non-trivial for $\alg{su}(2|1)$.

\section{The Leading Order Representation \label{leadingorder}}

\subsection{The $\alg{su}(2|1)$ Algebra}
For vanishing Yang-Mills coupling constant $g$, the $\alg{su}(2|1)$ generators act as the tensor product of single-site generators. The non-vanishing actions on a single site are:
\begin{align}
(\gen{Q}_a)_0 \state{\phi^b} &= \delta^b_a \state{\psi}, & \gen{R}^a_b \state{\phi^c} & =  \delta^c_b \state{\phi^a} - \frac{1}{2} \delta^a_b \state{\phi^c},\nln
(\gen{S}^a)_0 \state{\psi} & =   \state{\phi^a}, & \gen{D}_0 \state{\phi^a} & =  \state{\phi^a}, \nln
& & \gen{D}_0 \state{\psi} & = \frac{3}{2} \state{\psi}.
\end{align}
We have neglected the subscript $0$ for $\gen{R}^a_b$, since the compact $\alg{R}$ symmetry receives no quantum corrections.

The leading term of $\delta \gen{D}$, which is $\mathcal{O}(g^2)$, acts on adjacent sites of the spin chain as\footnote{For simplicity, we will write the action of generators on states of minimal length. The action on longer states is just given by the homogeneous sum over the length of the chain.} \cite{Beisert:2004ry}
\<
\gen{\delta D}_2 = 1 - \Pi,
\>
where $1$ is the identity operator, and $\Pi$ is the graded permutation operator. $\gen{\delta D}_2$ also equals the quadratic Casimir operator:
\<
\gen{\delta D}_2 = g_{ab} \gen{J}^a(1) \gen{J}^b(2).
\>
For larger sectors (where the Casimir has more than two distinct eigenvalues on two-site chains), this simple relation between the Hamiltonian and the quadratic Casimir is replaced by a relation involving the digamma function \cite{Lipatov:1997vu, Kotikov:2000pm, Kotikov:2001sc, Dolan:2001tt, Beisert:2003jj}.

\subsection{The $\alg{su}(2|1)$ Yangian Algebra}
The local symmetry generators $\gen{J}^A_0$ are given by the previous section and (\ref{jdef}). The Yangian generators $\gen{Y}^A_0$ then act as%
\<
\gen{Y}^A_0=  f^A{}_{C B} \sum_{i < j} \gen{J}^B_0(i) \gen{J}^C_0 (j).
\>
The adjoint transformation rule (\ref{adjoint}) is satisfied because of the Jacobi identity. The Serre relation follows from a straightforward generalization of the proof given for $\alg{su}(n)$ Yangians in \cite{Dolan:2004ps}.

To show that the $\gen{Y}^A_0$ commute with the one-loop dilatation generator, one can modify the arguments used in \cite{Dolan:2003uh} for the full $\alg{psu}(2,2|4)$ spin chain, or check by explicit computation that on a chain of length $2$,
\[
\comm{\gen{\delta D}_2}{\gen{Y}^A_0} = \gen{J}^A_0(1) - \gen{J}^A_0(2). \label{gauge}
\]
Since $\gen{\delta D}_2$ commutes with the $\gen{J}^A$, the commutator evaluated on longer chains is just the sum of this adjacent $2$-site commutator over the length of the chain. However, this yields a total chain derivative which vanishes on an infinite chain. Because chain derivatives vanish on periodic states, which correspond in the Yang-Mills theory to gauge-invariant traces of operators, we will use gauge transformations as a synonym for chain derivatives.

\section{The Next-to-Leading Order Corrections \label{oneloop}}

\subsection{The $\alg{su}(2|1)$ Algebra}

The one-loop correction to the supersymmetry generators are two-site operators. The non-vanishing actions are \cite{Beisert:2003ys, Zwiebel:2005er}
\<
(\gen{Q}_a)_2 \state{\phi^b \phi^c} \eq \frac{1}{4} \delta^b_a  (\state{\psi \phi^c}- \state{\phi^c \psi})  + \frac{1}{4} \delta^c_a  (\state{\phi^b \psi}- \state{\psi \phi^b} ),  \nln
(\gen{Q}_a)_2 \state{\phi^b \psi}  \eq \frac{1}{4} \delta^b_a \state{\psi \psi} , \nln
(\gen{Q}_a)_2 \state{\psi \phi^b }   \eq - \frac{1}{4} \delta^b_a \state{\psi \psi} , \nln
(\gen{S}^a)_2 \state{\psi \phi^c} \eq  \frac{1}{4} (\state{\phi^a \phi^c}  - \state{\phi^c \phi^a },  \nln
(\gen{S}^a)_2 \state{ \phi^c \psi} \eq  \frac{1}{4} ( \state{\phi^c \phi^a } - \state{\phi^a \phi^c} ),  \nln
(\gen{S}^a)_2 \state{\psi \psi}  \eq \frac{1}{4} ( \state{\phi^a \psi} - \state{ \psi \phi^a})  .
\> 

We can generalize the quadratic product for the one-loop dilatation generator to a compact expression for all of the one-loop generators, 
\<
\gen{J}^A_2 &=& \frac{1}{8} \left( d_{BC}{}^{A}-d^A{}_{BC} \right) \gen{J}_0^B(1) \gen{J}^C_0(2) + \frac{1}{8}  ((-1)^{(AA)}-1) \left( \gen{J}^A_0(1) \gen{J}^8_0(2) + \gen{J}^8_0(1) \gen{J}^A_0(2) \right) \notag \\ 
&&+ \frac{1}{2} g^{A8} g_{BC} \gen{J}^B_0(1) \gen{J}^C_0(2).
\>
Importantly, note that this expression is basis-dependent. Also, the first line vanishes for all of the bosonic generators, and the second line gives the dilatation generator.

At two-loops, the dilatation generator already has a lengthy expression. We use the expression in \cite{Beisert:2003ys} for the two-loop Hamiltonian of the $\alg{su}(2|3)$ sector, which can be applied directly here by letting the Latin indices represent two types of bosons and Greek indices represent the fermion. Again, there is also a more compact expression,
\<
\gen{\delta D}_4 & = & \, \left( \frac{1}{8} d_{CBA} - \frac{1}{8} (-1)^{AB} d_{BCA} \right) \gen{J}^A_0(1) \gen{J}^B_0 (2) \gen{J}^C_0(3) \notag \\
&& +\frac{1}{8} (g_{AB}-g_{BA}) \left(\gen{J}^8_0(1) \gen{J}^A_0(2) \gen{J}^B_0(3)+ \gen{J}^A_0(1) \gen{J}^B_0(2) \gen{J}^8_0(3) \right) \notag \\
& & - \frac{1}{2} \gen{J}^8_0(2) \gen{\delta D}_2(1,3) + \gen{\delta  D}_2(1,3) - 2 \, \gen{\delta D}_2.
\>
As above, this expression is basis dependent. The expression in \cite{Beisert:2003ys} is more general, as it includes possible similarity and gauge transformations, but they have no effect on the spectrum. 
\subsection{The $\alg{su}(2|1)$ Yangian Algebra}

We find that the next-to-leading order corrections to the Yangian generators can still be written in terms of the $\gen{J}^A$.
\<
\gen{Y}^A_2 &=&  \gen{Y}^A_{2\text{ non-local}} + \gen{Y}^A_{2\text{ local}}, \label{oneloopyangian} \\
\gen{Y}^A_{2\text{ non-local}} &=&    f^A{}_{C B} \left( \sum_{i < j-1 }  \gen{J}^B_2(i,i+1) \gen{J}^C_0 (j)+  \sum_{i < j }   \gen{J}^B_0(i) \gen{J}^C _2(j,j+1) \right), \\
\gen{Y}^A_{2\text{ local}} &=&     \sum_{i} f^A{}_{C B} \gen{J}^B_2(i,i+1) (\gen{J}^C_0(i+1) - \gen{J}^C_0(i))   \nln
  &+& \sum_{i} f^A{}_{C B} (\gen{J}^B_0(i)- \gen{J}^B_0(i+1)) \gen{J}^C_2(i,i+1)   \label{localyangian} \\
 &+&  \sum_i  \left( \alpha  \, \gen{Y}^A_0(i, i+1)   + \gamma \sum_{B=1}^8 ( g^{BA}- g^{AB}) \left( \gen{J}^B_0(i)-\gen{J}^B_0(i+1) \right) \right). \notag \>
It is interesting to note that if the minus signs in the first two lines of (\ref{localyangian}) were replaced by plus signs, it would just be $\gen{J}^A_2$ acting on the entire chain. Also, for $A=8$ the first two lines of (\ref{localyangian}) vanish. Note that the term proportional to $\gamma$ vanishes on infinite chains since it is a gauge transformation. We include it because the choice $\gamma=1$ simplifies the proof that the Serre relation is satisfied. However, when checking commutators of the $\gen{Y}^A$ with the $\gen{J}^B$ we set $\gamma=0$, since commutators of gauge transformations and local generators are gauge transformations. 
For vanishing $\alpha$ and $\gamma$, (\ref{localyangian}) is the sum over the chain of the two-site interaction
\begin{gather}
\gen{Y}^A_{2 \, \text{local}} =  \left(2 \,  f_{BC}{}^A + 3\, f^A{}_{CB} + 3 \, f_{CB}{}^A \right) \gen{J}^B_0(1) \gen{J}^C_0(2) \notag \\
+ \frac{1}{2} \, \sum_{B=1}^8 ( g^{BA}- g^{AB}) \left( \gen{J}^B_0(1) \gen{J}^8_0(2) - \gen{J}^8_0(1) \gen{J}^B_0(2) \right) - g^A{}_8 \gen{Y}^8_0(1,2). 
\end{gather}
Again, this is a basis dependent expression. We have verified that the Yangian $\alg{su}(2)$ generators restricted to the $\alg{su}(2)$ subsector agree with the expression found in \cite{Agarwal:2004sz} for $\alpha = \frac{1}{2}$. We cannot check with the $\alg{su}(1|1)$ Yangian presented in \cite{Agarwal:2005ed} since there the corrections to the Yangian generators are local, and bi-local corrections are essential when the $\alg{su}(1|1)$ sector's Yangian is embedded in a larger sector's Yangian.

\paragraph{Similarity Transformations.} As explained in \cite{Beisert:2003ys}, the local symmetry algebra is preserved by similarity transformations, which act as 
\<
\gen{J}^A \mapsto e^{\gen{U}} \gen{J}^A e^{-\gen{U}} = \gen{J}^A + \comm{\gen{U}}{ \gen{J}^A}+ \ldots, 
\>
and the Yangian symmetry is preserved also if we apply the same transformation to the $\gen{Y}^A$. In general, $\gen{U}$ could be any generator acting on the chain, but we require similarity transformations to preserve the coupling dependence that arises from Feynman diagrams, the manifest $R$-symmetry, and the even parity of the local symmetry generators. As a result, the $\mathcal{O}(g^2)$ contributions from $\gen{U}$ can be linear combination of only five possible two-site interactions. Two linear combinations of these, the identity and the one-loop dilatation generator, commute with all of the $\gen{J}^A_0$ and $\gen{Y}^A_0$. So there are three non-trivial independent similarity transformations. One linear combination preserves the form (\ref{oneloopyangian}), while the remaining two are not consistent with this form. It is intriguing  that the elegant expression found for the $\mathcal{O}(g^2)$ symmetry generators of the $\alg{psu}(1,1|2)$ sector (which contains the $\alg{su}(2|1)$ sector) in \cite{Zwiebel:2005er} gives the basis that we have used.

\paragraph{Adjoint Transformation.} We now verify that the adjoint transformation rule (\ref{adjoint}) is satisfied at next-to-leading order.  First, we just consider the commutator with $\gen{J}^8$, which beyond leading order is the anomalous part of the dilatation generator. Since corrections to each of the $\gen{Y}^A$ have the same dimension as at leading order, the vanishing commutator of the leading order $\gen{Y}^A$ with the one-loop dilatation generator implies that at next-to-leading order the Yangian charges transform properly under commutators with $\gen{J}^8$ \cite{Dolan:2004ys}. 

The same proof as at leading order shows that the term  of $\gen{Y}^B_2$ proportional to $\alpha$ transforms properly with respect to the $\gen{J}^A$. This term is needed (again with $\alpha = \frac{1}{2}$) for the commutator with the dilatation generator at $\mathcal{O}(g^4)$, but is not necessary for the $\gen{J}^A$ and $\gen{Y}^A$ to form a Yangian algebra at next-to-leading order. Of the remaining terms, those that involve products of generators far apart transform properly under commutation with the $\gen{J}^A$, as a straightforward generalization of the leading order proof would still work. However, we need to check terms in the commutator that involve one generator that intersects the other two generators. All of these terms involve only two adjacent sites of the chain, so it will be sufficient to examine this commutator on a chain with just two sites. Explicitly, we need to check the second of the following equalities:
\begin{eqnarray} \gcomm{\gen{J}^A}{\gen{Y}^B}_{2 \, \text{local}} & = & f^B{}_{D C}  \left(  \gcomm{\gen{J}^A_2(1,2)}{\gen{J}^C_0(1) \gen{J}^D_0(2)}  \right. \notag  \\
& &+     \gcomm{\gen{J}^A_0(1) + \gen{J}^A_0(2)}{ \gen{J}^C_2(1, 2) (\gen{J}^D_0(2) - \gen{J}^D_0(1))}  \notag \\
& & + \left.  \gcomm{\gen{J}^A_0(1) + \gen{J}^A_0(2)}{ (\gen{J}^C_0(1)- \gen{J}^C_0(2)) \gen{J}^D_2(1,2) } \right)  \notag \\
& = &  f^{AB}{}_C \gen{Y}^C_{2 \, \text{local}} + \text{terms that vanish on an infinite chain}. \end{eqnarray}
Canceling terms, we find that this is equivalent to
 \<
 0 &= & f^{B}{}_{DC}\left(\gcomm{\gen{J}^A_2(1,2)}{\gen{J}^C_0(1)} \gen{J}^D_0(1) - (-1)^{AC} \gen{J}^C_0(1) \gcomm{\gen{J}^A_2(1,2)}{\gen{J}^D_0(1)} \right)  - \text{parity}  \nln  & & -\frac{1}{8} \gcomm{\gen{J}^A_2(1,2)}{\gen{Y}^B_0(1,2)} + \text{terms that vanish on an infinite chain}. \label{equivalent}
 \>
Here `parity' just means interchange sites $1$ and $2$. Since $\gen{J}^A$ receives no quantum corrections for $A=5,\,6,\text{ or }7$, clearly Yangian generators transform properly with respect to these local symmetry generators. More generally, (\ref{equivalent}) simplifies to 
 \< 
&& m^{AB}{}_{C} \left( \gen{J}^C_0(1) - \gen{J}^C_0(2) \right) \label{adjointgauge}, 
 \>
where in our basis 
 \< 
m^{AB}{}_{C}&=& \frac{1}{8} ((-1)^{AA} -1) \left( \sum_{D=1}^8 g^{DA}  f^{BD}{}_C + 2 \, g^{B8} g^A{}_C + 2 \, g^8{}_Cg^{AB} \right)  - g^{A8} g^B{}_C. \nln 
 \>
The expression (\ref{adjointgauge})  is just a gauge transformation. Therefore, the Yangian charges transform properly under the local symmetry algebra.
 
\paragraph{Commutator with the Hamiltonian.} We have also verified that the dilatation generator commutes with this Yangian representation up to $\mathcal{O}(g^4)$. Again the commutator splits into local and bi-local parts. The local part includes terms where one of the generators intersects both of the others, as well as the commutator involving the last term in the expression for $\gen{Y}^B_2$. The rest of the commutator is bi-local.  The bi-local part vanishes because the dilatation generator commutes with the $\gen{J}^A$. The local piece involves three adjacent sites. Using \texttt{Mathematica}, we have checked this commutator on a chain of three-sites:
\<
\comm{\gen{\delta D}}{\gen{Y}^A}_{4\text{ local}} & = & \comm{\gen{\delta D}_2}{\gen{Y}^A_2}_{\text{local}} + \comm{\gen{\delta D}_4}{\gen{Y}^A_0}_{\text{ local}} \nln
&=& \gen{W}^A(2,3) - \gen{W}^A(1,2), \nln
\gen{W}^A(1,2) &=& 4\, d^A{}_{CB} \gen{J}^B_0(1)\gen{J}_0^C(2)+ 4 \, \gen{J}^A_0(1) + 4 \, \gen{J}^A(2) + 8\,  \gen{J}^A_2(1, 2).
\>
Since the difference of the $\gen{W}^A$'s is a gauge transformation, the $\gen{Y}^A$ commute with $\gen{\delta D}$ at $\mathcal{O}(g^4)$.

\subsection{The Serre Relation}
The Serre relation is satisfied up to a new type of term that vanishes on an infinite chain,
  \[ \begin{split}
 f^{[BC}{}_{E}\gcomm{\gen{Y}^{A \} }}{\gen{Y}^E}_2 &= -\frac{64}{3}  (-1)^{(EM)} f^{AK}{}_{D} f^B{}_{E}{}^{L} f^C{}_{F}{}^{M}  f_{KLM}\{ \gen{J}^D, \gen{J}^E, \gen{J}^F \}_2  \label{modifiedserre} \\
 & \quad + a^{ABC}{}_{DE} \left( \gen{J}^D_0(1)  \sum_{i=2}^{L}\gen{J}^E_0(i) + \gen{J}^D_0(L) \sum_{i=1}^{L-1}\gen{J}^E_0(i) \right).  \end{split}
 \]
We have not found simple expressions for the coefficients $a^{ABC}{}_{DE}$ (we have found their values in our basis), but this will not be necessary. To show that this last term vanishes on an infinite chain, we introduce the set of parity-odd two-site operators
\[
\gen{\tilde{Y}}^{AB}(i, j) = \gen{J}^A_0(i) \gen{J}^B_0(j) - \gen{J}^A_0(j) \gen{J}^B_0(i).
\]
It is always possible to find coefficients $c_{ABC}{}^{DE}$ such that 
\[
c_{ABC}{}^{DE} \gcomm{\gen{J}_0^A(i)}{\gen{\tilde{Y}}^{BC}(i, j)} = \gen{J}_0^D(i) \gen{J}_0^E(j).
\]
Using this, we write the extra term in (\ref{modifiedserre}) as
\[
\sum_{i<j} \tilde{a}^{ABC}{}_{DEF} \gcomm{\gen{J}^D_0(1)- \gen{J}^D_0(L)}{\gen{\tilde{Y}}^{EF}(i, j)},
\]
for some new coefficients $\tilde{a}$. The first term in the commutator vanishes on infinite chains. Therefore, satisfying (\ref{modifiedserre}) is equivalent to satisfying the Serre relations on infinite chains. It is simpler to check the Serre relation when the extra term is written as in (\ref{modifiedserre}).

We will now prove that (\ref{modifiedserre}) is satisfied. We have checked this equation on two and three-site chains, for $\gamma=1$. In fact, this is sufficient to guarantee that this relation holds for any length chain. Define\footnote{Note that $\gen{Z}^{ABC}(1, L)$ acts on all sites between and including sites $1$ and $L$.}
\[
\gen{Z}^{ABC}(1, L) =  f^{[BC}{}_{E}\gcomm{\gen{Y}^{A \} }}{\gen{Y}^E}_2+ \frac{64}{3}  (-1)^{(EM)} f^{AK}{}_{D} f^B{}_{E}{}^{L} f^C{}_{F}{}^{M}  f_{KLM}\{ \gen{J}^D, \gen{J}^E, \gen{J}^F \}_2,
\]
evaluated on a chain of length $L$, which we need to show equals the extra term in (\ref{modifiedserre}) for any $L$. Also, define $\gen{B}^{ABC}(1, L)$ to be the terms of  $\gen{Z}^{ABC}(1, L)$ simultaneously including generators acting on the first site and generators acting on the last site. 

The terms entering the Serre relation are local, bi-local, or tri-local.  Local terms are those that would appear on a chain of length two. The contribution from the cubic $\gen{J}$ term actually vanishes on two sites, but the $\gen{Y}$ commutators yield
\[
\gen{Z}^{ABC}(1,2) =   a^{ABC}{}_{DE} ( \gen{J}^D_0(1) \gen{J}^E_0(2)+\gen{J}^D_0(2) \gen{J}^E_0(1)),
\] 
which agrees with (\ref{modifiedserre}) for $L=2$.

Bi-local terms first appear for chains of length three. The bi-local terms from the cubic $\gen{J}$ expression are canceled by commutators of  the non-local part of $\gen{Y}_2$ with a twice-intersecting $\gen{Y}_0$ (not including terms where a nearest-neighbor $\gen{Y}_0$ acts on the same sites as a $\gen{J}_2$). The remaining bi-local pieces either include a local $\gen{Y}_2$ or a $\gen{J}_2$ intersecting a $\gen{Y}_0$ on two sites. For these terms, all that matters is that two-sites are adjacent, so that a $\gen{J}_2$ or $\gen{Y}_{2\text{ local}}$ act on them, and the location of the third site does not matter. Using this, we can determine the contribution from the bi-local terms just from a three-site \texttt{Mathematica} computation. For the bi-local part involving the boundaries of a chain of length $3$ we find
\<
\gen{B}^{ABC}(1, 3) =  a^{ABC}{}_{DE} \left( (\gen{J}_0^D(1)- \gen{J}_0^D(2)) \gen{J}_0^E(3) + (\gen{J}_0^D(3) - \gen{J}_0^D(2)) \gen{J}_0^E(1) \right). \label{boundary3} 
\>

The remaining terms are tri-local. The tri-local piece gives zero total contribution to the $\gen{Z}$ since the $\gen{J}$ satisfy the $\alg{su}(2|1)$ algebra exactly without gauge transformations. Therefore, the bi-local boundary contributions (\ref{boundary3}) for chains of length three are the only contributions to $\gen{B}^{ABC}$ for any chain of length greater than two, 
\<
\gen{B}^{ABC}(1, L) =  a^{ABC}{}_{DE} \left( (\gen{J}_0^D(1)- \gen{J}_0^D(2)) \gen{J}_0^E(L) + (\gen{J}_0^D(L) - \gen{J}_0^D(L-1)) \gen{J}_0^E(1) \right). \label{boundary} \nln
\>

Now we can finish the proof using induction. Assume that (\ref{modifiedserre}) is satisfied for chains of length $L$. Then split terms of $\gen{Z}^{ABC}(1,  L+1)$ into those that only act on the first $L$ sites, those that only act on the last $L$ sites, and those that act on both boundaries. However, we also need to subtract the terms that only act on the intersection of the first and last $L$ sites. Then using our assumption and substituting (\ref{boundary}) is sufficient: 
\<
\gen{Z}(1,  L+1)&=& \gen{Z}^{ABC}(1,  L)+ \gen{Z}^{ABC}(2,  L+1)- \gen{Z}^{ABC}(2,  L) + \gen{B}^{ABC}(1, L+1)  \nln
&=& a^{ABC}{}_{DE} \left( \gen{J}^D_0(1) \sum_{i=2}^{L}\gen{J}^E_0(i) + \gen{J}^D_0(L) \sum_{i=1}^{L-1}\gen{J}^E_0(i)+ \gen{J}^D_0(2) \sum_{i=3}^{L+1}\gen{J}^E_0(i) \right. \nln
&& \quad + \left. \gen{J}^D_0(L+1) \sum_{i=2}^{L}\gen{J}^E_0(i) -  \gen{J}^D_0(2) \sum_{i=3}^{L}\gen{J}^E_0(i) - \gen{J}^D_0(L) \sum_{i=2}^{L-1}\gen{J}^E_0(i) \right)   \nln
&& \quad +   \gen{B}^{ABC}(1, L+1) \nln
&=& a^{ABC}{}_{DE} \left(\gen{J}^D_0(1) \sum_{i=2}^{L}\gen{J}^E_0(i) + \gen{J}^D_0(2) \gen{J}^E_0(L+1) + \gen{J}^D_0(L) \gen{J}^E_0(1) \right. \nln
&& \quad +   \left. \gen{J}^D_0(L+1) \sum_{i=2}^{L}\gen{J}^E_0(i) \right) + \gen{B}^{ABC}(1, L+1)  \nln
&=& a^{ABC}{}_{DE} \left( \gen{J}^D_0(1)  \sum_{i=2}^{L+1}\gen{J}^E_0(i) + \gen{J}^D_0(L+1) \sum_{i=1}^{L}\gen{J}^E_0(i) \right), 
\>
in agreement with (\ref{modifiedserre}). The Serre relation is satisfied regardless of the value of $\alpha$ (the coefficient of $\gen{Y}^A_0(i, i+1)$ in the expression for $\gen{Y}^A_2$), but we do not have a simple explanation for this.

\section{Discussion \label{conclusion}}
We have constructed the next-to-leading order corrections to the $\alg{su}(2|1)$ sector Yangian, proving integrability at two-loops. Furthermore, these corrections are built in a simple way from the local symmetry generators. Perhaps the most important result is the generalization of the standard definition of Yangian symmetry to include this system. The local symmetry generators still transform as usual,
\[ 
\gcomm{\gen{J}^A}{\gen{J}^B} = f^{AB}{}_{C} \gen{J}^C \label{local2}. 
\]
However, the adjoint transformation rule of the Yangian charges is generalized to allow for gauge transformations. This had to be the case since at leading order the Hamiltonian only commutes with the Yangian generators up to gauge transformations, and the Hamiltonian becomes part of the symmetry algebra starting at next-to-leading order. On a chain of length $L$ the adjoint transformation rule is
\[\gcomm{\gen{J}^A}{\gen{Y}^B} = f^{AB}{}_{C} \gen{Y}^C + m^{AB}{}_C (\gen{\hat{J}}^C(\text{left}) - \gen{\hat{J}}^C(\text{right})). \label{adjoint2}
\]
The only essential part of the last term is that it is a boundary term. At next-to-leading order in this sector, $\gen{\hat{J}}^A(\text{left})$ is $\gen{J}^A_0(1)$ and $\gen{\hat{J}}^A(\text{right})$ is $\gen{J}^A_0(L)$. At higher orders we expect that the $\gen{\hat{J}}^A$ will at least involve higher order corrections to the $\gen{J}^A$ acting on the boundary sites and their immediate neighbors, and in principle they do not even need to be simply related to the $\gen{J}^A$.
Finally, the Serre relation is generalized to include commutators of boundary terms,
\< 
 f^{[BC}{}_{E}\gcomm{\gen{Y}^{A \} }}{\gen{Y}^E}  &=& h^2  (-1)^{(EM)} f^{AK}{}_{D} f^B{}_{E}{}^{L} f^C{}_{F}{}^{M}  f_{KLM}\{ \gen{J}^D, \gen{J}^E, \gen{J}^F \} \nln
&& \quad +  \tilde{a}^{ABC}{}_{DEF} \gcomm{\gen{\hat{J}}^D_0(\text{left})- \gen{\hat{J}}^D_0(\text{right})}{\gen{\tilde{Y}}^{EF}}\label{serre2} . 
 \>
Again, the only essential part of the new term is that $\gen{\hat{J}}^D_0(\text{left})- \gen{\hat{J}}^D_0(\text{right})$ is a boundary term.

A natural question is on the importance of the $\alg{su}(2|1)$ algebra for this construction. For instance, is it possible to repeat this construction for other $\alg{su}(n|m)$ algebras, with spins in the fundamental representation? If not, what is special about $\alg{su}(2|1)$? Also, it would be interesting to see how much of the structure we have found for the Yangian charges can be maintained at higher order in this sector. Are the local pieces of the higher order corrections still just given by the appropriate choice of signs for the products of overlapping symmetry generators, combined with multiples of the local pieces of lower-order Yangian charges? 

We have laid the foundation for generalizations to the $\alg{psu}(1,1|2)$ and $\alg{su}(2|3)$  sectors. New features would appear for these sectors, non-compactness for the former and length-changing generators for the latter. Also, for the $\alg{su}(2|3)$ sector the local symmetry generators only commute up to gauge transformations, so in this case we will have to generalize the defining Yangian algebra equations even further. We anticipate that the adjoint transformation rule will need to include commutators with boundary terms, and the Serre relation will need to include commutators with commutators of boundary terms.

While higher-order and larger sector Yangian computations will become impractical very rapidly, further study may yield new insights that simplify the construction of the Yangian and deepen our understanding of the integrability of these spin-chain models. Finally, the constraints imposed by constructing the Yangian algebra should be useful in finding, or even deriving, loop corrections to the Hamiltonian.

\appendix
\section{$\alg{su}(2|1)$ Basis}

The transformation between the two different bases for the local $\alg{su}(2|1)$ symmetry generators is given by
\begin{gather}
\gen{J}^1=\gen{S}^1 + \gen{Q}^1, \quad \gen{J}^2=i (\gen{S}^1 - \gen{Q}^1),  \quad \gen{J}^3 =   \gen{S}^2 + \gen{Q}^2,  \quad \gen{J}^4 =  i (\gen{S}^2 - \gen{Q}^2), \notag \\
\gen{J}^5 = \gen{R}^1_2 + \gen{R}^2_1, \quad \gen{J}^6 = i (\gen{R}^{1}_2 - \gen{R}^2 _1), \quad \gen{J}^7 = \gen{R}^1_1 - \gen{R}^2_2, \quad \gen{J}^8 = 2 \gen{D}_0 - \gen{L}+ \gen{\delta D}. \label{jdef}
\end{gather}
$\gen{L}$ is the length generator, which commutes with all of the other symmetry generators (there are no length-changing symmetries in this sector).

Up to permutations of the indices, the only non-vanishing structure constants are:
\<
f^{117} &= &f^{135} =  f^{227} = f^{236} = f^{245} = -2, \nln
 f^{118} &=& f^{146} =  f^{228} = f^{337} = f^{338} =  f^{447} = f^{448}=2, \nln
f^{567} &= &4 i.
\>

The non-vanishing components of $g$ are 
\<
g_{12} = g_{34} =\frac{i}{2}, \quad g_{21} = g_{43} = - \frac{i}{2}, \quad g_{55} = g_{66}=g_{77}=- \frac{1}{2} \quad g_{88}= \frac{1}{2}.
\>

Up to permutations of the indices, the only non-vanishing symmetric invariant tensor components are:
\begin{align}
d^{127} &= d^{136} =  d^{145} = d^{246}  = -2 i,& d^{235} &= d^{347} =  2 i,\nln
d^{128} & = d^{348} = - 6 i,& d^{558} &=  d^{668} = d^{788} = 4, \nln
d^{888} &= -12. & &
\end{align}
%

\subsection*{Acknowledgments}

I thank my advisor, Niklas Beisert, and Abhishek Agarwal, Denis Erkal, and Chiara Nappi for valuable discussions and comments. I am grateful to the Albert Einstein Institute Potsdam, where this work was completed. This article is partly based upon work supported by a National Science Foundation Graduate Research Fellowship. Any opinions, findings and conclusions or recommendations expressed in this material are those of the author and do not necessarily reflect the views of the National Science Foundation.


\end{document}